\documentclass[journal,10pt,twocolumn]{IEEEtran}
\usepackage{epsfig}
\usepackage{graphicx}
\usepackage{subfigure}
\usepackage{threeparttable}

\usepackage{color}
\usepackage{amssymb}
\usepackage{makeidx}
\usepackage{amsmath}
\usepackage{graphicx}
\usepackage{epsf}
\usepackage{epsfig}
\usepackage{psfig}
\usepackage{ccaption}
\usepackage{array}
\usepackage{tabularx}
\usepackage{multirow}
\usepackage{epsfig}
\usepackage{cite}
\usepackage{procedure,algorithm,algorithmic}
\usepackage{authblk}

\begin{document}

\title{\LARGE Survey on UWB Positioning System: Sampling methods}
\author[*]{Sheng Jiang}
\author[**]{Matthew Trinkle}
\author[***]{Hongyuan Wang}
\affil[*]{Department of Computer Science, Huazhong University of Scientist and Technology, Wuhan, China, 430074}
\affil[**]{School of Electrical and Electronic Engineering,The University of Adelaide, Adelaide, Australia, 08831}
\affil[***]{Department of Electrical Information, Huazhong University of Scientist and Technology, Wuhan, China, 430074}

\maketitle

\begin{abstract}
Millimeter-accuracy Ultra-Wideband (UWB) positioning systems using the Time Difference Of Arrival (TDOA) algorithm are able to be utilized in military and many other important applications. Previous research on UWB positioning system has achieved up to mm or sub-mm accuracy. However, one bottleneck in UWB system is at sampling high resolution UWB signals, as well as high resolution timing information. In this paper, UWB positioning systems are surveyed and we focus on sampling methods for handling UWB signals. Among different sampling methods, one traditional way is the sequential sampling method, which is not a real time sampling method and blocks UWB positioning system to achieve higher precision. Another way is by applying Compressed Sensing (CS) to UWB system for achieving sub-mm positioning accuracy. In this paper, we compare different TDOA-based UWB systems with different sampling methods. In particular, several CS-UWB algorithms for UWB signal reconstruction are compared in terms of positioning accuracy. Simulation results in 2D and 3D experiments demonstrate performance of different algorithms including typical BCS, OMP and BP algorithms. CS-UWB is also compared with UWB positioning system based on the sequential sampling method.
\end{abstract}

\section{Introduction}
High accuracy indoor positioning system is becoming an increasingly critical in many wireless applications where location is of high importance, e.g., personnel tracking and monitoring in office environments, tracking of patients and high value assets in hospitals, and indoor Robert navigation. One typical application is the remotely controlled Robert surgery system which requires real time wireless mm or sub-mm accuracy. The higher the accuracy, the greater value the positioning system to both users and suppliers. The ultra-wideband (UWB) technology can be used to track a moving target within mm level accuracy since the transmitted signal has an extremely short duration (typically in the order of sub-nanosecond), which can provide ultra high resolution timing information. Angle Of Arrival (AOA) and Time Difference Of Arrival (TDOA) algorithms can be exploited to achieve good 3D positioning accuracy. However, one big challenge in UWB positioning systems is how to obtain high resolution timing information.

The previous UWB positioning system \cite{rws}\cite{maf} using  sequential sampling method can achieve sub-cm accuracy. Fig. 1 illustrates the GPS-similar 3D indoor Robert UWB positioning system for the application of remotely controlled Robert surgery.
A transmitter, called {\em the tag}, periodically sends out
ultra short duration pulses (about 300-ps duration) at a certain frequency which is known at receiver sides.
Surrounding the tag, four receivers, called {\em base stations (BS)}, can receive the transmitted UWB pulses. Due to geometrical difference, the pulse arrival time at difference base stations are also different. TDOA algorithm is then exploited to calculate the position of the tag.
In order to obtain high resolution timing information, we adopted a sequential sampling method to acquire UWB signals. The basic idea is to utilize a sampling clock at base station which has a small offset compared with the pulse repetition frequency for UWB signal acquisition. Then one period of UWB signal can be composed by many periods of original signals since the received UWB signal is regarded to be repetitive at a fixed pulse repetition frequency\cite{rws}. However, the sequential sub-sampler suffers from many problems to achieve higher accuracy. One of the biggest problem is that it will enlarge the geometrical error since it is not a real time signal acquisition method. In order to achieve an ultra-high positioning accuracy in real time with the TDOA algorithm, we have to utilize a real time sampling method. However, ultra-high sampling rate ADCs for UWB signal acquisition are either commercially unavailable or unaffordable expensive.

Compressed sensing (CS) theory \cite{Candes2006} can be applied to UWB systems to alleviate the sampling problem. The previous research has applied CS into UWB systems for UWB signal acquisition  with a low sampling rate \cite{cissadc}\cite{pimrc}\cite{ciss_1}\cite{yang_thesis}\cite{ICC2010_9}\cite{EURASIP_13}\cite{CISS2011_16}. A joint space-time Bayesian Compressed Sensing (BCS) algorithm \cite{eurosip}\cite{DSP}\cite{yang_thesis} can be utilized for UWB positioning systems but it is complicated and does not exploit pulse template redundancy in UWB signals.
Other related studies in \cite{UWBchannel} focus on UWB signal acquisition using traditional CS algorithms.
Authors in \cite{dcs1}\cite{dcs2} developed a general joint signal reconstruction algorithm which may be utilized for joint UWB signal reconstruction in UWB positioning systems. However, those algorithms are for a general case which do not consider any template redundancy in UWB signals.
In \cite{2010pulse}, the CS algorithm has been modified for reconstructing pulse stream signal. In \cite{network_position}, CS theory is applied to positioning systems. Other most recent development on UWB positions are in \cite{2013_1}\cite{2013_2}\cite{2013_3}. However, it is performed in the frequency domain. Another disadvantage for CS is its computational complexity, which blocks it applying into more applications such as UWB wireless communications\cite{2013_4}\cite{dyspan_4}\cite{TWC_6}\cite{TSG_15}\cite{CISS2011_16}\cite{EL_17}. And, computation acceleration optimization for CS and BCS have been developed to speedup computation \cite{fccm_3}\cite{SAAHPC_7}\cite{SAAHPC2010_10}\cite{SAAHPC20102_11}\cite{parallel_21} so that UWB systems can take advantages of CS algorithms.

\begin{figure}
  \centering
  \includegraphics[scale=0.45]{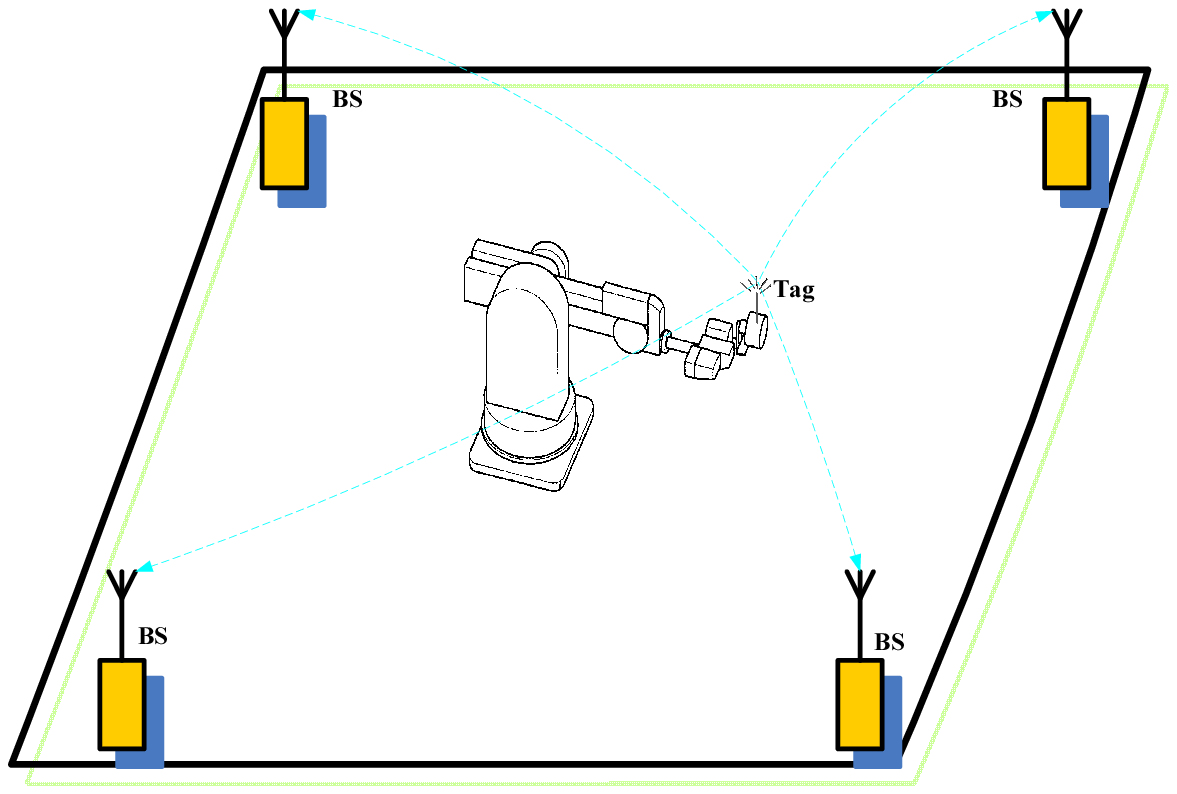}
  \caption{A typical UWB positioning system}\label{fig:sys}
\end{figure}

This paper surveys recent progress of UWB positioning systems. One traditional sampling method is to utilize sequential sampling. While the newest method is about CS-based UWB positioning system using low sampling rate ADCs and the TDOA algorithm to achieve sub-mm accuracy\cite{yang_thesis}\cite{DSP}\cite{ciss_1} . We note that there exist plenty of redundancy or \emph{a priori} information in the UWB positioning system. The received UWB signal is a combination of the transmitted UWB pulse, where the pulse shape is called as the template \emph{a priori} information. And the received UWB signals among base stations are similar in space, where there exist the spatial\emph{ a priori} information. The CS-based UWB positioning system is to utilize these redundancies. By introducing template and spatial  \emph{a priori} information, the CS-UWB algorithm can further lower down the sampling rate and improve the capability of de-noise for UWB signal reconstruction compared with other traditional CS algorithms. The optimized CS-UWB positioning algorithm can quickly compute the position of the tag for fast target tracking.

Numerical simulation results investigate the performance of the CS-UWB positioning algorithm for 1D, 2D, and 3D UWB positioning error. The tested UWB echo signals are drawn from  the IEEE802.11b UWB standards with frequency-dependent propagation channels\cite{IEEE}.
Compared with the traditional OMP, BCS and BP algorithms, the CS-UWB algorithm can reconstruct the UWB signal at the lowest sampling rate. And the CS-UWB algorithm can achieve the best positioning performance in the 1D experiment. In 2D and 3D experiments,
simulation results show that CS-UWB positioning algorithm can achieve
the best accuracy than the traditional sequential sampling method. Therefore, BCS based UWB positioning system as shown in \cite{yang_thesis}\cite{DSP}\cite{EURASIP_13} has better performance compared with other similar systems.

The remainder of this paper is organized as follows. The signal model of the CS-based UWB
positioning system using the sequential sampling method and CS scheme are formulated in Section \ref{sec:model}, \ref{sec:model_sq} and \ref{sec:model_cs}. We first introduce and analyze the BCS algorithm in Section \ref{sec:BCS}. The template and spatial \emph{a priori} information in the BCS algorithm are mainly introduced in Section \ref{sec:CS-UWB}. Simulation results of comparing different algorithms and sampling methods are demonstrated in Section \ref{sec:simu}. Finally, Section \ref{sec:con} concludes the paper.


\section{Signal Model}\label{sec:model}
In this section, we will introduce the UWB signal model. Then we model the acquired UWB signal using the sequential sampling method based on previous work\cite{rws}. Finally, the CS-based UWB positioning system model is presented.

\subsection{General signal model}
The UWB pulse is repetitively transmitted at a fixed frequency. The received UWB signal at one pulse repetition period is defined as {\em a frame}. Then in the continuous time domain, a frame of UWB signal, $s(t)$, received at a base station through multi-path channels can be expressed as:
\begin{eqnarray}\label{eq:pulse}
s(t)&=&\sum_{m=1}^M a_{m}p(t-t_{m})\nonumber\\
&=& \sum_{m=1}^M a_{m}\exp{\left(-\frac{(t-t_m)^2}{2\sigma^2}\right)},
\end{eqnarray}
where  $p(t)$ is the transmitted Gaussian pulse; $\sigma$ represents the width of the pulse; $m$ is the number of resolvable propagation
paths; $a_{m}$ is the amplitude attenuation of the signal along the
$m$-th path and $t_{m}$ is the time delay of the $m$-th path. It is clear that the received signal is actually a combination of the transmitted Gaussian pulse with different delays and amplitudes. Since the propagation channels are frequency dependent, the received pulse may have some distortion on the pulse shape. But we assume the received pulse is very similar to the transmitted pulse\cite{uwbbook}, which can be approximately modeled by Eq. (\ref{eq:pulse}).
This pulse shape information in the received UWB signal is named as template \emph{a priori} information.
At the same time, the received UWB signals at different base stations are also very similar, which can be described as spatial \emph{a priori} information.

From the received signal at different base stations, the pulse arrival time can be detected. We denote the pulse arrival at $j$-th and $i$-th base stations as $t^{BSj}$ and $t^{BSi}$ respectively. Then the time difference $\tau_{ji}$ is obtained as:
\begin{eqnarray}\label{eq:td}
\tau_{ji}=t^{BSj}-t^{BSi}=c*(d_{j}-d_{i})
\end{eqnarray}
where $d_{j}$ represents the distance between the tag and $j$-th base station, and $d_{j}$ represents the distance between the tag and $i$-th base station. $c$ is the propagation speed of microwave. The time difference $\tau_{ji}$ implies the distance difference information. In this paper, we assume the first arrival pulse peak indicates the pulse arrival time in a LOS environment\cite{uwbbook}, which has the largest signal amplitude\cite{rws}. When multiple time difference information are collected from several base stations, the position of the tag can be calculated by using the TDOA algorithm \cite{rws}\cite{maf}.

However, the TDOA algorithm requires a high sampling rate to get the fine timing information for high positioning precision. In previous work\cite{maf}, we adopt a sequential sampling method for UWB signal acquisition.

\section{UWB positioning system based on sequential sampling method}\label{sec:model_sq}
The sequential sampling method is to utilize a small relative frequency offset between the tag and the receiver to sequentially extend the received UWB pulses. Assume the pulse repetition frequency is $f_p$ and the sampling frequency at a base station is $f_s$. So we have the equivalent sampling rate,
\begin{eqnarray}
f_{eq}=|\frac{1}{\frac{1}{f_p}-\frac{1}{f_s}}|=|\frac{f_sf_p}{f_s-f_p}|
\end{eqnarray}
It is observed that the equivalent sampling rate $f_{eq}$ is proportional to the difference of $f_s$ and $f_p$. Considering the frequency draft between $f_s$ and $f_p$, the extended received signal and the original signal can be expressed as:
\begin{eqnarray}
s(t)&=&s(t+\frac{1}{f_p})\\
r(n)&=&s(t+ n\frac{1}{f_s\pm\Delta f})
\end{eqnarray}
where $r(n)$ is the digitalized signal using the sampling frequency $f_s$ with a relative draft frequency $\Delta f$. And $s(t)$ is the original UWB signal. In the time domain, comparing $s(t)$ and $r(n)$, the time scale is actually extended by a scale $K_r$,
\begin{eqnarray}
K_r=\frac{f_p}{|f_s\pm\Delta f-f_p|}=K\pm\Delta k
\end{eqnarray}
where $K$ is the expected scale number without the draft frequency. Consequently, the pulse arrival time detected from the signal $r(n)$ should be:
\begin{eqnarray}
t_{real}^{BSi}&=&\frac{t^{BSi}_{r(n)}}{K_r}=\frac{t^{BSi}}{K\pm\Delta k}\\
d^{BSi}&=&\frac{t_{real}^{BSi}}{c}=\frac{t^{BSi}}{c(K\pm\Delta k)}
\end{eqnarray}
where $t^{BSi}_{r(n)}$ is the pulse arrival time detected based on the time-extended signal $r(n)$ at the $i$-th BS. $d^{BSi}$ is the distance from the tag to the $i$-th BS.

Obviously, the calculated distance $d^{BSi}$ is proportional to the extension scale $K_r$, which is effected by the draft frequency $\Delta K$. This error will also cause a serious geometrical dilution. Normally, in the experiment, this $\Delta k$ is about several percentages of $K$\cite{rws}. However, the estimated pulse arrival time $t_{real}^{BSi}$ will deviate from the true value when the tag is moving away from the center of base stations.
In order to overcome this problem, we have to adopt a real time high resolution sampling method. We apply CS theory to UWB systems to achieve real time sampling but using low sampling rate ADCs.


\section{UWB positioning system based on compressed sensing}\label{sec:model_cs}
\begin{figure}
  \centering
  \includegraphics[scale=0.5]{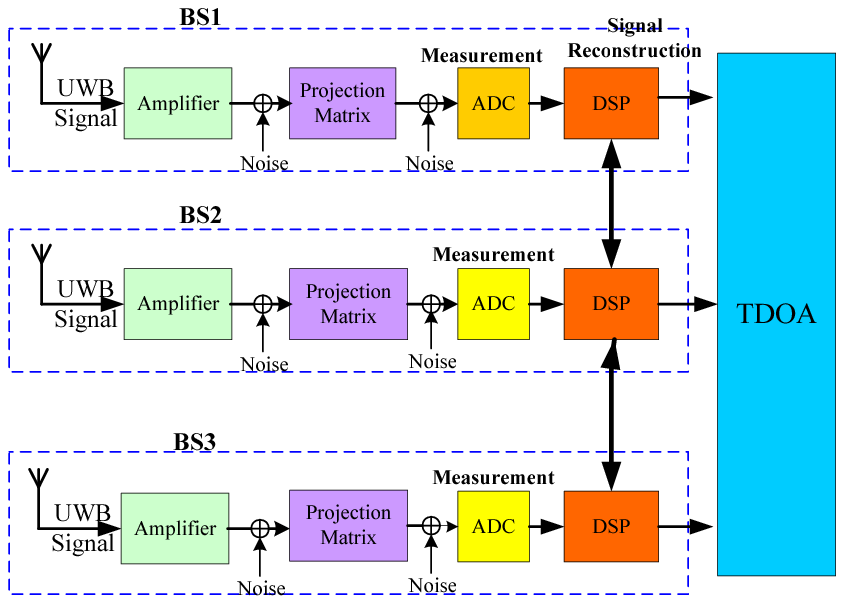}
  \caption{Compressed sensing-based UWB positioning system}\label{fig:sysCS}
\end{figure}

Fig. \ref{fig:sysCS} shows the structure of the CS-based UWB positioning system. The amplified received UWB after the antenna is fed into the analog matrix to mix the original UWB signal, which can be obtained using low sampling rate ADCs to yield a small amount of measurements. A possible analog hardware implementation scheme of the projection
matrix can be found in \cite{cissadc}. Then the measurement $\boldsymbol{y}$ at a base station is given by:
\begin{eqnarray} \label{eq:cs}
\boldsymbol{y}=\boldsymbol{\Phi} (\boldsymbol{s+n_1}) +
\boldsymbol{n_2}=\boldsymbol{\Phi} \boldsymbol{s} +\boldsymbol{\epsilon}
\end{eqnarray}
where $\boldsymbol{\Phi}$ is a projection matrix $\boldsymbol{\Phi}$, ($\boldsymbol{\Phi} \in R^{M \times N}$) at a base station. $\boldsymbol{y}$ is a $M$-dimensional measurement vector. The UWB signal $\boldsymbol{s}$ is a N-dimensional vector, representing a frame of digitized signal. The additive noise $n$ on the signal may be from propagation channel, hardware devices, or coupled space noise. And the noise $n_2$ is added on  the measurements, which may be from projection matrix and digitalization error from limited quantity effects of ADCs. The overall noise is $\boldsymbol{\epsilon}=\Phi (n_1+n_2)$. For mathematical convenience, we model the noise $\boldsymbol{\epsilon}$ as a additive white Gaussian distribution with zero-mean and variance $\beta$.

The UWB signal $\boldsymbol{s}$ can be reconstructed from $\boldsymbol{y}$
using CS algorithms. Note that the received UWB signal is well known sparse, i.e. most elements in $\boldsymbol{s}$ are zero so that the amount of the measurements can be  dramatically smaller than the amount of signal at a time interval, i.e., $M<<N$. Then we can collect measurements at a low sampling rate to reconstruct the high resolution UWB signal. From the reconstructed signal, the pulse arrival time can be determined and thus the position of the tag.

\subsection{Bayesian Compressed Sensing}\label{sec:BCS}
In this section, we first introduce the framework of the BCS algorithm. Then the key step, maximizing the log-likelihood function, is analyzed for UWB signal reconstruction.

\subsection{Bayesian compressed sensing framework}

The BCS algorithm builds a Bayesian regression approach to reconstruct the original signal from measurements\cite{bcs}. Starting from the Gaussian distributed noise model, it implies a multivariate Gaussian distribution, which is given by
\begin{eqnarray}
P(\boldsymbol{y}|\boldsymbol{s},\beta)=(2\pi)^{-N/2}\beta^{-N}\exp\{-\frac{\|\boldsymbol{y}-\boldsymbol{\Phi} \boldsymbol{s}\|^2}{2\beta^2}\}
\end{eqnarray}

The key of the BCS algorithm is to impose an exponential distribution on  each signal element, which is given by
\begin{eqnarray}
P(\boldsymbol{s|\alpha}) &=&\prod_{i=1}^N(\frac{\alpha_i}{2 \pi})^{1/2} \exp\{-(s_i)^2 \frac{\alpha_i}{2}\} \nonumber\\
&\sim&
\mathcal{N}(\boldsymbol{s_{}}|0,\boldsymbol{{(\alpha_{})^{-2}}})
\end{eqnarray}

The signal obeys a multidimensional Gaussian distribution, which is given by

\begin{eqnarray} \label{eq:pw}
P(\boldsymbol{s|y,\alpha},\beta) &=& \frac{P(\boldsymbol{y|s},{\beta} )P(\boldsymbol{s|\alpha})}{P(\boldsymbol{y|\alpha},{\beta})}\nonumber\\
&\sim&
\mathcal{N}(\boldsymbol{s}|\boldsymbol{\mu},\boldsymbol{\Sigma})
\end{eqnarray}
where $\boldsymbol{A} = \mbox{diag}(\boldsymbol{\alpha^i})$. The
covariance and the mean of the signal are given by
\begin{eqnarray}
\boldsymbol{\Sigma}=\left(\beta^{-2}
(\boldsymbol{\Phi})^{T}\boldsymbol{\Phi} +
\boldsymbol{A}\right)^{-1}
\end{eqnarray}
and
\begin{eqnarray}\label{eq:es_sig}
\boldsymbol{\mu}=\beta^{-2}\Sigma
\boldsymbol{(\Phi)^{T}y}
\end{eqnarray}

Therefore, the mean of the distribution of the signal vector is regarded as the estimation of the signal, $\boldsymbol{\hat s}$, and the covariance can be viewed as the "error bar".
The reconstructed signal vector can be written as:
\begin{eqnarray} \label{eq:signal}
\boldsymbol{\hat s}
&=& \beta^{-2}\Sigma(\Phi)^T y \nonumber\\
&=&
(\boldsymbol{\Phi}^{T}\boldsymbol{\Phi} + \beta ^2
\boldsymbol{A})^{-1} (\boldsymbol{\Phi})^{T} \boldsymbol{y}
\end{eqnarray}

Note that the hyperparameter matrix $\boldsymbol{A}$ plays a key role. Without the hyperparameters, BCS algorithm will degrade to the least square method. In order to estimate the hyperparameters by maximizing the marginal log-likelihood function, which is given by:
\begin{eqnarray}\label{eq:max}
\boldsymbol{L(\alpha)}= \log P(\boldsymbol{y|\alpha,\beta})=
\int P(\boldsymbol{y|s},\beta)P(\boldsymbol{s|\alpha})ds
\end{eqnarray}

Expectation-Maximization (EM) and incremental optimization methods in \cite{frvm} can be utilized to maximize the objective function for estimating $\boldsymbol{A}$. The incremental optimization method is similar to the OMP algorithm, which is much faster than EM. So that we adopt the incremental optimization method \cite{frvm} to optimize the objective function for UWB signal reconstruction. More details are given in Appendix \ref{app:bcs} for completeness.
In order to modify the BCS algorithm for UWB signal reconstruction, we need to analyze the objective function $\boldsymbol{L(\alpha)}$.

\section{Compressed Sensing UWB Signal Reconstruction}\label{sec:CS-UWB}
In order to improve the performance of reconstructing UWB signals, we need to exploit helpful redundant or \emph{a priori }information in UWB signals. There are two types of \emph{a priori } information in the UWB positioning system:
\begin{itemize}
  \item Template \emph{a priori }information. At one base station, the received UWB signal can be regarded as a combination of the transmitted UWB pulse with different amplitudes and delays. Whin one frame, the nonzero elements are clustered; it is impossible to have an ��isolated�� nonzero element which means that its neighboring elements are both zero. Moreover, as long as the pulse peak is obtained, the amplitudes of neighboring elements can also be approximately estimated according to the transmitted pulse shape.
  \item Spatial \emph{a priori }information. The transmitted UWB pulse is intercepted at
      multiple base stations, thus incurring spatial  \emph{a priori }information; The received UWB signals among different base stations are very similar. Therefore, we can combine the information of multiple base
      stations to exploit the spatial redundancy.
\end{itemize}
More details about CS-UWB signal reconstruction by modifying the BCS algorithm for UWB positioning systems can be find in \cite{yang_thesis}\cite{DSP}\cite{EURASIP_13}\cite{RWS_14}.

\subsection{Compressed Sensing UWB Positioning Algorithm}\label{sec:algo}

The TDOA algorithm is performed with the CS-UWB positioning algorithm in parallel\cite{yang_thesis}\cite{DSP}\cite{RWS_14}. The iterative TDOA algorithm is detailed in Appendix \ref{app:TDOA}. Traditionally, the pulse arrival time cannot be obtained until the UWB echo signal is fully reconstructed. However, both CS and TDOA algorithms are computationally expensive. The CS-UWB positioning algorithm can be utilized for fast target tracking. At each iteration, the pulse arrival time is computed and forwarded to the TDOA algorithm for computing the tag position
while the UWB signal reconstruction procedure is still ongoing. This is based on the fact that the acquisition of the pulse arrival time does not always require a perfect signal recovery. And the result form the TDOA algorithm is set as an initial for the next calculation. This method can help the TDOA algorithm converge quickly as long as the initial position is close to the true position.

the CS-UWB positioning algorithm for the UWB positioning system is detailed in \cite{yang_thesis}\cite{DSP}\cite{RWS_14}.  In CS signal reconstruction, the convergence condition depends on whether the signal is well reconstructed. However, for UWB positioning system, the convergency condition is depending on the detected pulse arrive time, which is given by:
\begin{eqnarray}
|t_{i+1}-t_{i}|<\varepsilon
\end{eqnarray}
where $t_{i+1}$ and $t_{i}$ are defined as the pulse arrival time obtained at two consecutive iterations. $\varepsilon$ is a small value depending on the final target positioning accuracy. Note that the noise level $\beta$ on the measurements is known so that the noise will not effect the signal reconstruction, which is ignored in the algorithm.

\section{Simulation Results}\label{sec:simu}
Numerical simulations are conducted to investigate the performance
of the CS-based UWB positioning system. In order to achieve mm accuracy, we adopt a 300 pico-second width Gaussian pulse as the transmitted signal \cite{rws}. The tested lossy UWB propagation channels are drawn from the experimental IEEE 802.15.4a UWB standards \cite{IEEE}. The received UWB pulse has more or less shape distortion due to frequency dependent loss on propagation channels. And the channel characteristics are unknown to the receivers.

We first demonstrate the performance of the CS-UWB algorithm for UWB signal reconstruction compared with the traditional OMP, BCS, and BP. Then 1D positioning error is illustrated using different CS reconstruction algorithms with different amount of measurements. Finally, in 2D and 3D scenarios,  the CS-based UWB positioning system is compared with the UWB positioning system using the traditional sequential sampling method.

\subsection{1D UWB Signal reconstruction}
In the CS-based UWB positioning system, the high resolution $N$-dimensional UWB signal vector can be indirectly reconstructed from the $M$-dimensional measurement vector. Since $M<N$, the sampling rate to obtain UWB signal is greatly reduced. We defined the reduction ratio of the sampling rate as:
\begin{eqnarray} \label{eq:adc}
R_r=\frac{M}{N}=\frac{f_M}{f_N},
\end{eqnarray}
where in one frame of the received UWB signal. There have a digitized $N$-dimensional signal vector and a $M$-dimensional measurement vector. And $f_N$ represents the sampling rate to acquire the $N$-dimensional signal vector and $f_M$ is denoted as the sampling rate to collect the measurement vector.
Reduction ratio measures the reduction of the sampling rate by applying the CS theory compared with the direct signal acquisition. And we measure the quality of the
reconstructed signal in terms of the reconstruction
percentage, which is defined as
\begin{eqnarray}
P_{re}=1-\frac{\|\mathbf{s}-\hat{\mathbf{s}}\|_2 } { \|\mathbf{s}\|_2},
\end{eqnarray}
where $\mathbf{s}$ is the true signal and $\hat{\mathbf{s}}$ is
the reconstructed signal.

\begin{figure}
  \centering
  \includegraphics[scale=0.55]{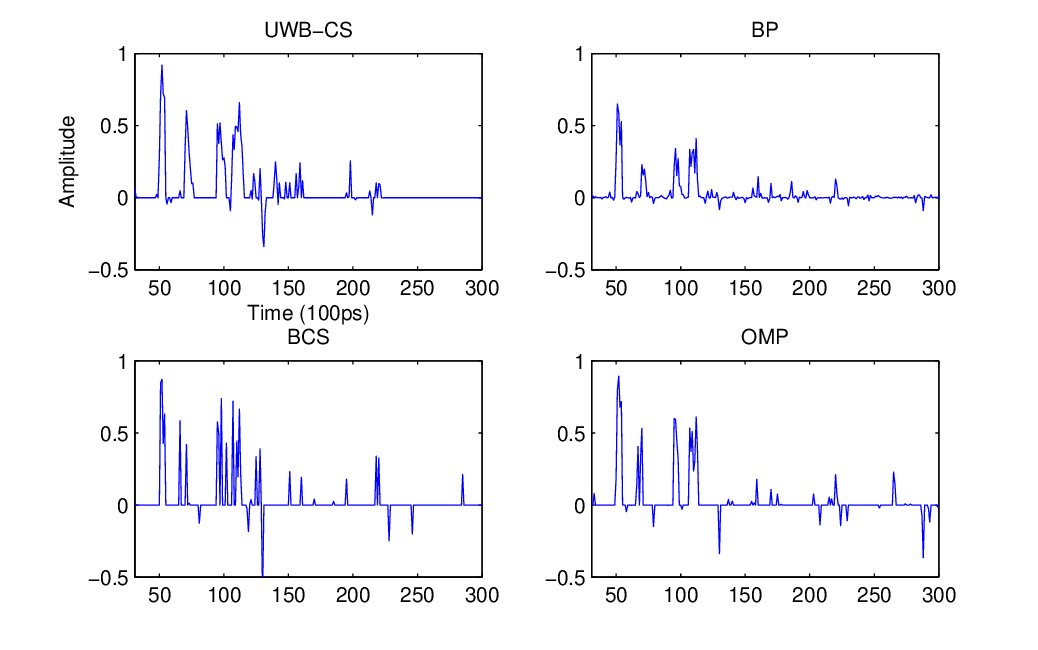}
  \caption{Signal reconstruction performance in time domain}\label{fig:sigtime}
\end{figure}

Fig. \ref{fig:sigtime} compares the reconstructed UWB signal in time domain using the CS-UWB algorithm and the traditional OMP, BCS and BP algorithms. All algorithms utilize the same amount of measurements ($R_r=0.15$) and $SNR\approx10dB$. The CS-UWB algorithm can achieve 48.2\% reconstruction percentage while BP can only approaches to 28.8\%, BCS 3.8\% and OMP 8.8\%. It is observed that the CS-UWB algorithm can recover the main pulse and correctly find the pulse arrival time with only 15\% of sampling rate of direct signal acquisition. In other words, the CS-UWB algorithm has the best performance in time domain compared with the traditional OMP, BCS and BP algorithms.

\begin{figure}
  \centering
  \includegraphics[scale=0.5]{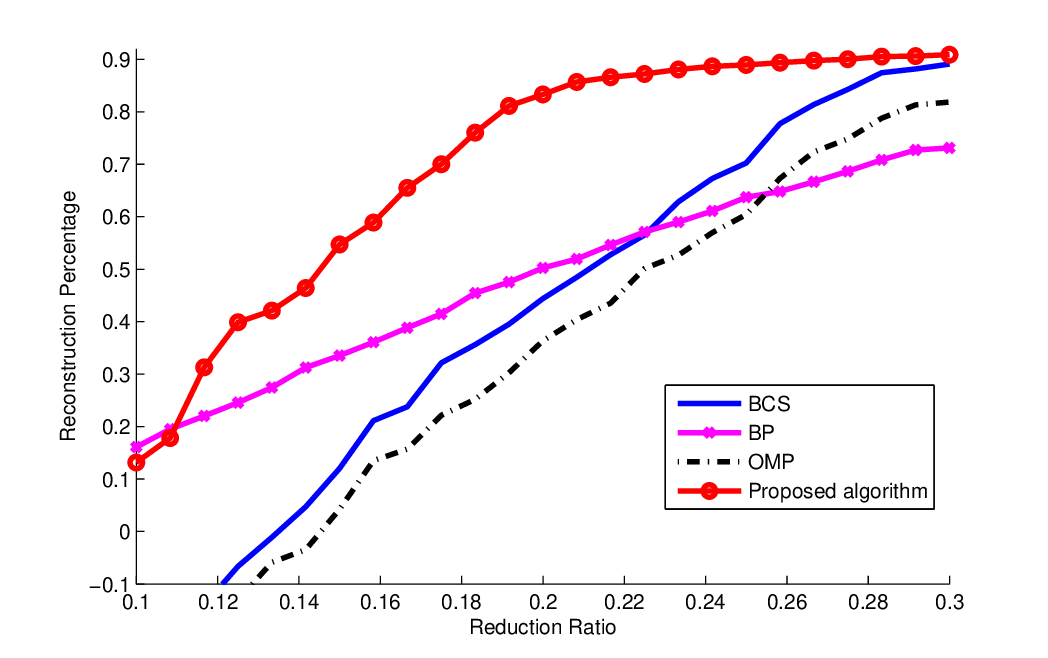}
  \caption{Signal reconstruction performance}\label{fig:sigrecon}
\end{figure}

Fig. \ref{fig:sigrecon} shows signal reconstruction performance with different sampling rate using different algorithms. The reduction rate varies from 0.1 to 0.3. With the growth of the sampling rate, the quality of the reconstructed signal increases for the CS-UWB, BP, BCS, and OMP algorithms. Note that the signal reconstruction percentage using the CS-UWB algorithm is much higher than other three algorithms at the same sampling rate. For instance, when the 0.21 reduction rate is 0.21, the CS-UWB algorithm can achieve to 81\% of reconstruction percentage while other three algorithms cannot approach more than 50\%.
Therefore, the CS-UWB algorithm still has the best performance for different sampling rates for UWB signal reconstruction.

\begin{figure}
  \centering
  \includegraphics[scale=0.5]{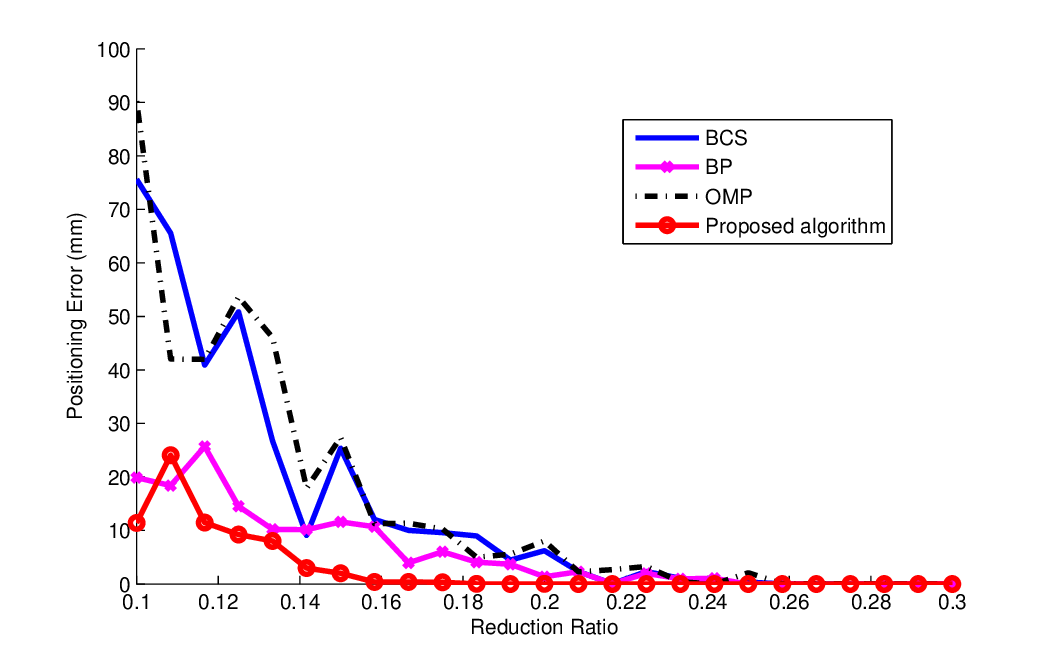}
  \caption{1D positioning performance}\label{fig:1derr}
\end{figure}

Fig. \ref{fig:1derr} illustrates 1D positioning performance using different algorithms. The error is calculated based on the difference between the true pulse peak and the pulse peak in the reconstructed UWB signal. If we utilize the reconstructed UWB signal for positioning purpose, the CS-UWB algorithm has the best positioning performance compare with the
OMP, BP and BCS algorithms. With the growth of the compression rate, the 1D positioning errors using all algorithms are all dramatically decreased since the reconstructed UWB signal are close to the true signal with more measurements. However, at the same compression ratio, the CS-UWB algorithm has the best positioning accuracy. And CS-UWB algorithm requires the lowest compression ratio compared with OMP, BP and BCS.

\subsection{2D UWB positioning performance}
The dominating factors in the UWB positioning system using the sequential sampling method are essentially different
from the CS-based UWB positioning system using CS-UWB positioning algorithm.
The sequential sampling method is the bottleneck to achieve a very high positioning accuracy in the UWB positioning system\cite{rws}\cite{maf} because it is not a real time signal acquisition method. We simulate 2D positioning performance using the CS-UWB positioning algorithm and the sequential sampling method \cite{rws}\cite{maf} for UWB positioning systems. Three base stations are utilized and the tested UWB echo signals have a certain level noise (SNR$\approx$10dB).

\begin{figure}
  \centering
  \includegraphics[scale=0.45]{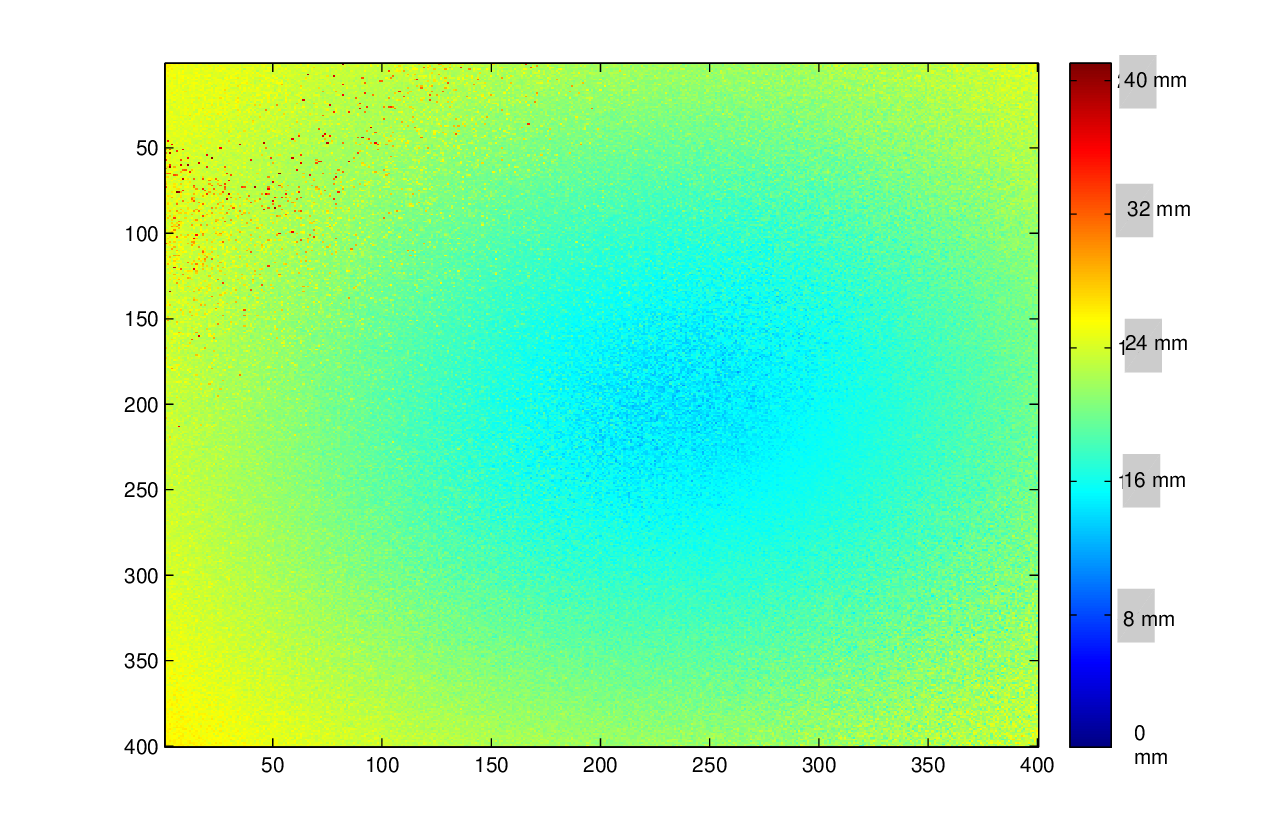}
  \caption{2D positioning error using sequential sampling method}\label{fig:2dsq}
\end{figure}

\begin{figure}
  \centering
  \includegraphics[scale=0.45]{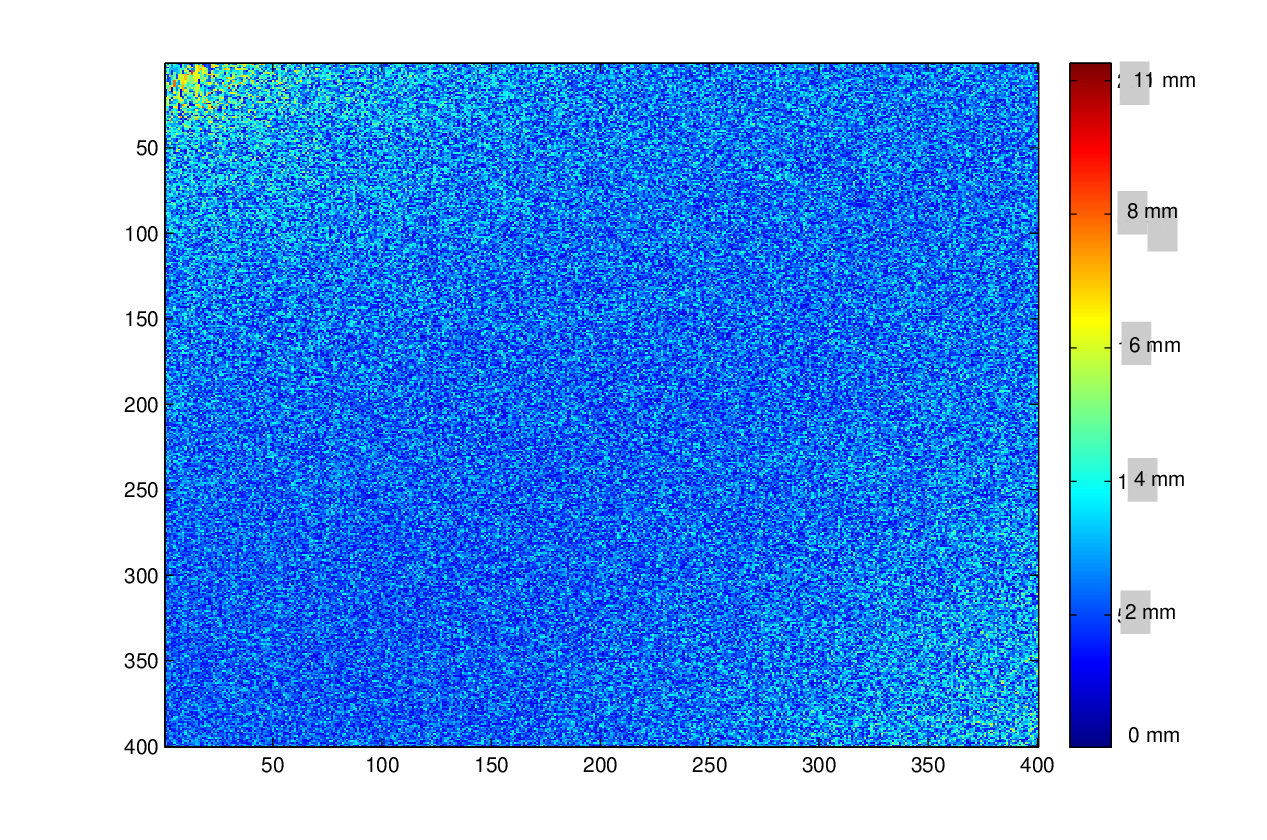}
  \caption{2D positioning error using the UWB-CS algorithm}\label{fig:2dcs}
\end{figure}

Fig. \ref{fig:2dsq} and Fig. \ref{fig:2dcs} show the 2D positioning error in a square area using the sequential sampling method and the UWB-CS algorithm.  It is observed that the error using CS scheme is even distributed but the error using sequential sampling method has a high accuracy area. The worse error accuracy in CS-based 2D UWB positioning system is about 10mm while the worse error is more than 40mm using the sequential sampling method. It is well known that there exist geometry error dilution in the TDOA algorithm. Unfortunately, UWB positioning system using the sequential sampling method can degrade this effect since the error of the pulse arrival time can not be accurately detected.

\subsection{3D UWB positioning performance}

We investigate 3D positioning performance using the CS-UWB positioning algorithm and the sequential sampling method \cite{rws}\cite{maf} for UWB positioning systems. The simulation is performed in a
5m$\times$5m$\times$4m room, where four base stations are placed at
the following positions: (0, 0,170), (4000, 0, 1855), (4410, 4435, 2860) and
(0, 4545, 3260) (in millimeters).
The noise is added to the original signals (SNR$\approx$10dB). The 3D UWB positioning error is based on the received UWB signals at base stations using the sequential sampling method and the CS-UWB positioning algorithms.

\begin{figure}
  \centering
  \includegraphics[scale=0.45]{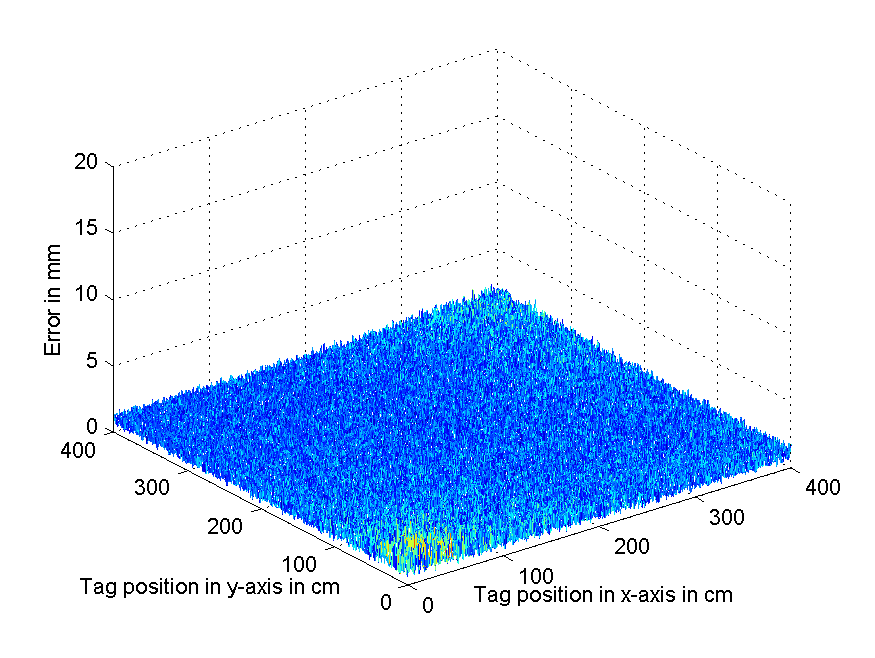}
  \caption{3D CS-based positioning system performance using the UWB-CS algorithm}\label{fig:3dcs}
\end{figure}

\begin{figure}
  \centering
  \includegraphics[scale=0.45]{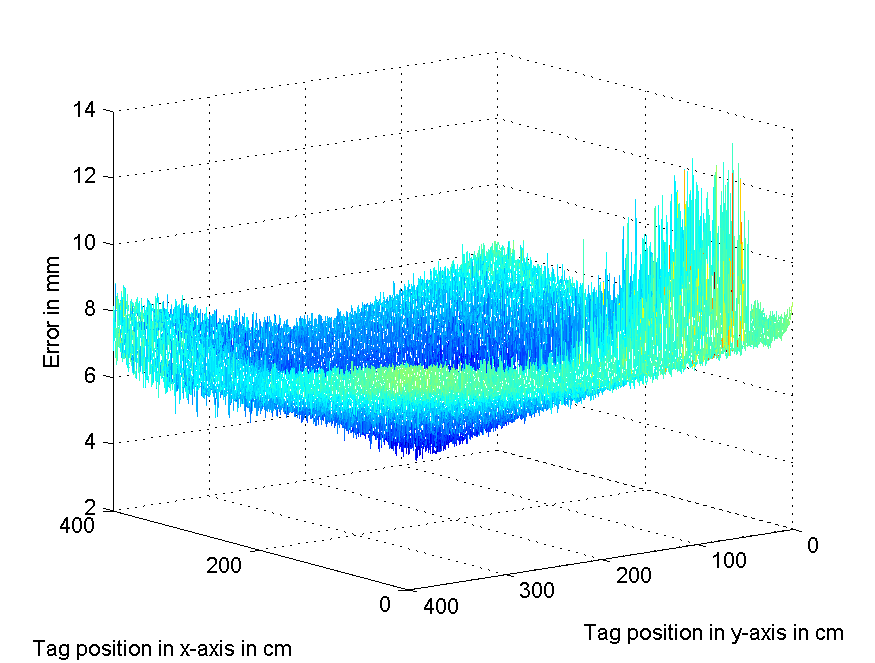}
  \caption{3D positioning system performance using sequential sampling method}\label{fig:3dsq}
\end{figure}

Fig. \ref{fig:3dcs} and Fig. \ref{fig:3dsq} show the performance comparison of the UWB positioning system using the CS-UWB positioning algorithm and the sequential sampling method. It is seen that the positioning accuracy is significantly improved by using the CS-UWB positioning algorithm in the CS-based UWB positioning system. The mean of error using CS-based sampling method is 0.92mm in Fig.\ref{fig:3dcs}; while it is about 6.45mm by using the sequential sampling method Fig. \ref{fig:3dsq}.
Moreover, Fig.\ref{fig:3dcs} shows that the error is unevenly distributed. The minimum error is achieved when the tag is at the geometrical center point, where the distance
difference is zero ($\tau_ji =0$). When the tag is moving close to the base
station and the distance differences are significantly increasing, the error will be
dramatically enlarged. However, when we apply CS theory into the UWB positioning system, the error is evenly distributed, as demonstrated in Fig. \ref{fig:3dsq}. The standard error variance
using the sequential sampling method as shown in Fig. \ref{fig:3dcs} is much larger than that using the scheme in Fig. \ref{fig:3dsq}. Besides improving the positioning accuracy, other advantages of using CS-UWB positioning algorithm in CS-based UWB positioning system include real time and high speed. And also note that the CS-based UWB positioning system using the CS-UWB positioning algorithm will be a breakthrough, which has a potential to achieve much higher accuracy.
Therefore, the positioning performance using the CS-UWB positioning algorithm in the CS-based UWB positioning system can be significantly improved compared with the system using the traditional sequential sampling method.
And the CS-UWB positioning algorithm can utilize \emph{a priori} information in CS-based UWB positioning system can obtain a high timing information but using a low compression ratio and low sampling rate ADCs to achieve a very high positioning accuracy.

\section{Conclusions}\label{sec:con}
In this paper, we survey UWB positioning systems with different sampling methods. The CS-based positioning system can achieve mm accuracy by using low sampling rate ADCs. Simulation results
compare performance of different UWB positioning systems, in which CS-based UWB positioning system is able to achieve significantly higher accuracy than the sequential sampling based UWB positioning system. Therefore, BCS based UWB positioning system as shown in \cite{yang_thesis}\cite{DSP}\cite{EURASIP_13} has better performance compared with other similar systems.

\appendices
\section{Fast maximizing the objective function}\label{app:bcs}

The key of the BCS algorithm is to optimize the objective function.
The marginal log-likelihood function is expanded to,
\begin{eqnarray} \label{eq:maxalphas}
\log p(\boldsymbol{y|\alpha},\beta) & = &
\log \int P(\boldsymbol{y|u},\beta)P(\boldsymbol{u|\alpha})d\boldsymbol{u}\nonumber\\
&=& -\frac{1}{2}(N log 2\pi + log |E| + y^T E^{-1}y)\nonumber\\
&=& L_1(\alpha_{-j}) + l_1(\alpha_j)
\end{eqnarray}
where
\begin{eqnarray}
E &=& \beta^2 I + \Phi A^{-1}\Phi^T\nonumber\\
 &=& \beta^2I + \sum_{k\neq j}\alpha_k^{-1} \phi_k \phi_k^{-1} + \alpha_j^{-1}\phi_j \phi_j \nonumber\\
&=& E_{-j} + \alpha_j^{-1}\phi_j \phi_j
\end{eqnarray}
and
\begin{eqnarray}
L_1(\alpha_{-i}) = -\frac{1}{2}(N log 2\pi + log |E_{-j}| + y^T E^{-1}_{-j} y)\\
l_1(\alpha_j) = \frac{1}{2}(log \alpha_j - log (\alpha_j + g_j) + \frac{h_j^2}{\alpha_j + g_j}
\end{eqnarray}
The quantities $g_j$ and $h_j$ are defined as
\begin{eqnarray}
g_j &=& \phi_j^T E_{-j}^{-1} \phi_j \label{eq:g_j}\\
h_j &=& \phi_j^T E_{-j}^{-1} y \label{eq:h_j}\\
E_{-j} &=& \beta^2I + \sum_{k\neq j}\alpha_k^{-1} \phi_k \phi_k^{-1}
\end{eqnarray}

By maximizing the term $l_1(\alpha_j)$, the optimal $\alpha_j$ is given by
\begin{equation}\label{eq:maxl_1}
\alpha_j =
\begin{cases}
 \frac{h_j^2}{g_j^2-h_j}, &\text{if $g_j^2 - h_j >0$;} \longrightarrow s_j \neq 0\\
 \infty , &\text{otherwise.} \longrightarrow s_j = 0
\end{cases}
\end{equation}

\section{TDOA algorithm}\label{app:TDOA}

Let ($x_i,y_i,z_i$), $i = 1,2,...I$ be the
known position coordinates of the $i$-th base station. And let ($x_t,y_t,z_t$) be the unknown tag location. The distance from the $i$-th base station to the tag
is denoted by $D_i$. Then, between the
$1$st and $i$-th base station, the difference of the pulse arrival time $\tau_{1i}$ can be obtained from the received UWB signal using Eq. (\ref{eq:td}). And the range difference ${D_{1i}}$ is given by
\begin{eqnarray}\label{timecal}
\tau_{1i} = cD_{1i}, \quad i=2,3,...I,
\end{eqnarray}
and
\begin{eqnarray}
D_{1i} = \sqrt{(x_1-x_t)^2+(y_1-y_t)^2+(z_1-z_t)^2} \nonumber\\
       - \sqrt{(x_i-x_t)^2+(y_i-y_t)^2+(z_i-z_t)^2},
\end{eqnarray}
where $c$ is the propagation speed. Taking the derivative on both sides of the
equation, we have
\begin{eqnarray}
d D_{1i} = \frac{(x_i-x_t)dx_t+(y_i-y_t)dy_t+(z_i-z_t)dz_t}{\sqrt{(x_i-x_t)^2+(y_i-y_t)^2+(z_i-z_t)^2}}\nonumber\\
 -
 \frac{(x_1-x_t)dx_t+(y_1-y_t)dy_t+(z_1-z_t)dz_t}{\sqrt{(x_1-x_t)^2+(y_1-y_t)^2+(z_1-z_t)^2}}.
 \end{eqnarray}
For the $i$-th base station with respect to the $1$st base station, the matrix is given by
\begin{eqnarray}\label{eq:tdoa}
\begin{pmatrix}
dD_{12} \\
dD_{13} \\
\vdots \\
dD_{1i}
\end{pmatrix}
=
\begin{pmatrix}
\alpha_{11} & \alpha_{12} & \alpha_{13} \\
\alpha_{21} & \alpha_{22} & \alpha_{23} \\
\vdots &  \vdots & \vdots \\
\alpha_{i1} & \alpha_{i2} & \alpha_{i3} \\
\end{pmatrix}
\begin{pmatrix}
dx_t \\
dy_t \\
dz_t
\end{pmatrix},
\end{eqnarray}
where, we have
\begin{eqnarray}
\alpha_{i1}=\frac{x_1-x_t}{D_1} - \frac{x_i-x_t}{D_i} \label{eq:a1}\\
\alpha_{i2}=\frac{y_1-y_t}{D_1} - \frac{y_i-y_t}{D_i} \label{eq:a2}\\
\alpha_{i3}=\frac{z_1-z_t}{D_1} - \frac{z_i-z_t}{D_i} \label{eq:a3}.
\end{eqnarray}
TDOA computation starts with an initial guess position of the tag, ($x_t,y_t,z_t$). By iteratively updating and solving Eq. (\ref{eq:tdoa}), TDOA will gradually converge to the true position. The computation continues until the error is below a certain threshold, which is given by
\begin{eqnarray}
err = \sqrt{(dx_t)^2+ (dy_t)^2+(dz_t)^2}.
\end{eqnarray}

\end{document}